\documentclass[11pt,a4paper]{article}
\pdfoutput=1
\usepackage{jheppub} 

\usepackage[T1]{fontenc} 

\usepackage{amsmath}
\usepackage{amssymb}
\usepackage{graphicx}
\usepackage{hhline}
\usepackage{mwe}

\usepackage{textcomp}
\makeatletter
\newcommand{\bea}{\begin{eqnarray}}
\newcommand{\eea}{\end{eqnarray}}

\newcommand{\pa}{\partial}

\newcommand{\be}{\begin{equation}}
\newcommand{\ee}{\end{equation}}
\newtheorem{lemma}{Lemma}

\title{\boldmath BPS soliton-impurity models and supersymmetry}

\author[a]{C. Adam}
\author[b,c,1]{Jose M. Queiruga,\note{Corresponding author.}}
\author[d]{A. Wereszczynski}

\affiliation[a]{Departamento de F\'isica de Part\'iculas, Universidad de Santiago de Compostela and Instituto Galego de F\'isica de Altas Enerxias (IGFAE) E-15782 Santiago de Compostela, Spain}
\affiliation[b]{Institute for Theoretical Physics, Karlsruhe Institute
of Technology (KIT), 76131 Karlsruhe, Germany}
\affiliation[c]{Institute for Nuclear Physics, Karlsruhe Institute of Technology (KIT), Hermann-von-Helmholtz-Platz 1,
D-76344 Eggenstein-Leopoldshafen, Germany}
\affiliation[d]{Institute of Physics, Jagiellonian University, Lojasiewicza 11, Krak\'{o}w, Poland}

\emailAdd{jose.queiruga@kit.edu}

\abstract{
We find  supersymmetric extensions of the half-BPS soliton-impurity models in (1+1) dimensions which preserve half of the $\mathcal{N}=1$ supersymmetry. This is related to the fact that in the bosonic sector (i.e., the half-BPS soliton-impurity model), only one soliton (for example, the kink) is a BPS configuration which solves the pertinent Bogomolnyi equation and saturates the topological energy bound. On the other hand, the topological charge conjugate state (the antikink) is not a BPS solution. This means that it obeys the full Euler-Lagrange equation and does not saturate the topological energy bound. 

The supersymmetric approach also allows us to construct half-BPS soliton-impurity models in (2+1) dimensions. Concretely,
in the case of the $CP^1$ model, its BPS impurity generalisation preserves one-quarter of the $\mathcal{N}=2$ SUSY, while for the Abelian 
Higgs model at critical coupling both impurity generalisations preserving one-quarter (the case of a new, so-called Higgs impurity) as well 
as one-half of the $\mathcal{N}=2$ SUSY (the case of the previously known magnetic impurity) 
are possible.

We also discuss a possible relation between the BPS $CP^1$-impurity model and the Dzyaloshinskii-Moriya interaction energy. }

\begin{document}
\maketitle
\flushbottom


\section{Introduction}
\label{intro}
It is widely known that the BPS (self-dual) solitons in (1+1) dimensions are 
intimately related to the existence of a $\mathcal{N}=1$ supersymmetric extension 
of the bosonic model. Here, by a BPS solution we understand a static solution of the so-called Bogomolnyi equations (which are of lower order than the Euler-Lagrange (EL) equations), which
saturates a pertinent topological energy bound. This guarantees the topological stability of the
solution for topologically nontrivial configurations (kinks). Furthermore, the BPS sector is equivalent to the zero pressure sector, as the Bogomolnyi equation is just the zero pressure condition. 

Supersymmetry provides a systematic tool for the derivation and analysis of such Bogomolnyi
equations. Indeed, they can be obtained from the $\mathcal{N}=1$ supersymmetric transformations of the fermions. Then, the self-dual sector (supporting kinks with topological charge $Q=1$) is invariant under one-half of the supersymmetry. Of course, the same happens with the anti self-dual sector. 

Furthermore, the model allows for a central extension,  where the central charge is the difference of the values of the superpotential at asymptotical values of the fields (vacua). The superpotential is related to the potential of the bosonic part of the theory by a target space differential equation.
Hence, for topologically nontrivial solutions (BPS solitons), the central charge takes a nonzero value.  

All these properties concern not only a scalar field theory with the standard kinetic term and an arbitrary (at least two vacua) potential but can be generalized to a quite arbitrary target space \cite{adam-santamaria} (multi-field and curved target space) as well as to models with nonstandard derivative terms \cite{k-model-1}-\cite{k-susy-7}. In fact, very recently it has been proved that the BPS property is shared even by theories with higher derivatives and, therefore, is a generic feature of {\it all} translational invariant scalar field Lagrangians in (1+1) dimensions \cite{higher-der}. Of course, this is at odds with higher dimensional models, where only very few models enjoy the BPS property. 

One could ask the obvious question of how to break the BPS-ness in (1+1) dimensions. This is possible but requires a quite drastic modification of the action, i.e., the addition of a term which breaks the translational invariance of the model, that is, an impurity (defect). Then, typically, no Bogomolnyi equations exist and solitons are solutions of the full EL equations. However, it has been found that there is a very special coupling of the defect which preserves  {\it one-half} of the BPS-ness \cite{higher-der}. This means that a kink (or antikink) is a BPS solution (solving a Bogomolnyi equation and saturating the pertinent bound) while its topological charge conjugated partner, i.e., the antikink (kink) does not have this property. 
 
 In the present work, we want to understand these half-BPS soliton-impurity standard scalar theories in (1+1) dimensions from a supersymmetric point of view. In particular, we will show that the Bogomolnyi equation again emerges via a supersymmetry transformation of fermions, leading to an invariance of the BPS sector under one-half of the SUSY. Similarly, we will obtain a central charge extension which, however, possesses only one nontrivial supercharge. Also the fermionic and bosonic zero modes coincide. This further explains the existence of the generalized translational symmetry of the BPS soliton. Finally, our approach allows for a derivation of a whole family of impurity deformed models which preserve one-half of the BPS-ness (and in the limit of the vanishing impurity reduce to the original scalar soliton BPS model). All such extensions preserve $1/2$ of the original $\mathcal{N}=1$ supersymmetry.
 
 \hspace*{0.2cm}
 
As $\mathcal{N}=1$ supersymmetry in (1+1) and (2+1) dimensions have basically the same structure, all our findings can be generalized to the (2+1) case. This gives us a chance to understand half-BPS preserving impurities in a unified way. Using these results, we construct half-BPS impurity extensions for the baby BPS Skyrme model. The Lagrangian with the impurity presents the same type of couplings as the scalar model in (1+1) dimensions. The only difference is the particular form of the topological current used in the construction, which for (1+1) gives the usual kinetic term while for the baby BPS Skyrme model is just the topological degree current. As a result, we get a model which preserves $1/4$ of $\mathcal{N}=2$ sypersymmetry. 

Interestingly, such a partially BPS preserving coupling of the impurity to a BPS model resembles in many aspects the partially BPS Abelian Higgs model at critical coupling with a magnetic impurity \cite{vortices}. In both theories, only half of the solitons enjoy the BPS property (are solutions of the pertinent Bogomolnyi equations) while the other half obey the full EL equations. Hence, only the former ones saturate the topological bound. Moreover, the impurity enters the Bogomolnyi equations of the original (no impurity) model as an inhomogenous term. Finally, the action requires the appearance of a coupling between a 'topological object' (the topological density or the magnetic field, respectively) and the impurity. 
 
Even more interestingly, we find the half-BPS preserving coupling of the impurity to the $CP^1$ model. In this case, the original Bogomolnyi equations, i.e., the Cauchy-Riemann (or anti Cauchy-Riemann) equations, are replaced by their non-homogeneous versions where the inhomogeneity is just the impurity. 
This implies the complete solvability in the BPS sector (which hosts half of the solitons of the original model).  
This result will enable us to introduce another impurity-Abelian Higgs model where  one-half of the BPS-ness is preserved. This is a different construction than the original one presented in \cite{vortices}. 
 
  \hspace*{0.2cm}

The last comment concerns our terminology. All  impurity models presented here are theories where one-half of the solitons are still BPS objects, so frequently we call them half-BPS soliton-impurity models. However, for the sake of simplicity sometimes we call them just BPS soliton-impurity models.
A related but different issue is the amount of SUSY preserved by the SUSY extensions of the impurity models.  Concretely, in the case of $\mathcal{N}=1$ SUSY always $1/2$ of the SUSY is preserved by the impurity BPS models, whereas in the $\mathcal{N}=2$ case both the preservation of $1/4$ or of $1/2$ of the supersymmetry are possible. We shall always denote these by $1/4$ SUSY and $1/2$ SUSY, respectively.


\section{The BPS preserving impurity in the scalar model in (1+1) dim}

\subsection{The BPS property from supersymmetry}
\label{BPS}

We will focus on $\mathcal{N}=1$ SUSY in $d=1+1$ dimensions. The modified Lagrangian with impurity $\sigma$ preserving one-half of the BPS property of the original model (without impurity) has the following form \cite{higher-der}
\be
\mathcal{L}=\frac{1}{2}\pa_\mu\phi\pa^\mu\phi-U-2 \sigma \sqrt{U}-\sqrt{2}\sigma \phi_x-\sigma^2. \label{sec:intro:imp}
\ee
Here $U$ is a (at least) two vacuum potential and $\phi_x\equiv \partial_x \phi$. 
We will give an explanation, based on a possible SUSY breaking, for the preservation of the BPS property. The first two terms in (\ref{sec:intro:imp}) have a simple SUSY extension
\be
\mathcal{L}_{0}=\frac{1}{4}\int d^2 \theta D^\alpha \Phi D_\alpha \Phi+\int d^2\theta \,W(\Phi) \vert= \frac{1}{2}\pa_\mu \phi\pa^\mu\phi+\frac{1}{2}F^2+W_\phi(\phi)F
\ee
where $\vert$ means setting $\psi_\alpha=0$. $\Phi$ is a scalar superfield and $D_\alpha$ is a superderivative, whose components are
\bea
\Phi&=&\phi+\theta^\alpha \psi_\alpha-\theta^2 F,\\
D_\alpha&=&\frac{\pa}{\pa \theta^\alpha}+i\theta^\beta \pa_{\alpha\beta}.
\eea
Since for (\ref{sec:intro:imp}) only the kink (or the antikink) are BPS solutions of the model, one should not expect to have a superfield formulation of the full model with impurity. If this were the case, then the existence of a BPS kink solution would imply the existence of the antikink and vice versa. Let us consider the following term
\be
\mathcal{L}_{impurity}^1=\sqrt{2}\phi_x \sigma.\label{sec:intro:imps}
\ee
Taking into account the supersymmetric transformations of the fields
\bea
\delta\phi&=&-\epsilon^\alpha \psi_\alpha,\\
\delta\psi_\beta&=&-\epsilon^\beta\left(C_{\alpha\beta} F+i\pa_{\alpha\beta}\phi\right),\label{sec:intro:tpsi}\\
\delta F&=&-\epsilon^\alpha i\pa_\alpha^{\,\,\beta}\psi_\beta, \label{sec:intro:tF}
\eea
we have
 \be
 \delta \mathcal{L}_{impurity}^1=-\sqrt{2}\sigma (\epsilon^\alpha \psi_{\alpha,x}).\label{sec:intro:sigma1}
 \ee
 If the impurity $\sigma$ is trivial ($\sigma_x=0$), then (\ref{sec:intro:sigma1}) is a total derivative and SUSY is preserved. If $\sigma_x\neq0$, this term has to be compensated in order to preserve (a part of) supersymmetry. The transformation (\ref{sec:intro:tF}) suggests the addition of the following term
  \be
\mathcal{L}_{impurity}^2=\sqrt{2}\sigma F.\label{sec:intro:impF}
 \ee
 The combination of (\ref{sec:intro:imps}) and (\ref{sec:intro:impF}) gives the following SUSY (static) transformations
 \bea
\frac{1}{\sqrt{2}} \delta ( \mathcal{L}_{impurity}^1+ \mathcal{L}_{impurity}^2)&=&-2\sigma \epsilon^2\psi_{2,x}-\sigma\epsilon^1\psi_{2,t}+\sigma\epsilon^2\psi_{1,t}, \label{sec:intro:varp}\\
\frac{1}{\sqrt{2}} \delta ( \mathcal{L}_{impurity}^1- \mathcal{L}_{impurity}^2)&=&- 2\sigma \epsilon^1\psi_{1,x}-\sigma\epsilon^2\psi_{1,t}+\sigma\epsilon^1\psi_{2,t}.\label{sec:intro:varm}
 \eea
 
 The conclusion is as follows: if we add $\mathcal{L}_{impurity}^1+ \mathcal{L}_{impurity}^2$ to the Lagrangian then $1/2$ SUSY is preserved (provided that $\epsilon^2=0$), while the combination $ \mathcal{L}_{impurity}^1- \mathcal{L}_{impurity}^2$ preserves $1/2$ SUSY (if $\epsilon^1=0$). Phrased differently, the addition of the impurity breaks explicitly one-half of the supersymmetry generators (one real Grassmann degree for $\mathcal{N}=1$ in $d=1+1$). We have therefore
 \bea
  \delta ( \mathcal{L}_{impurity}^1+ \mathcal{L}_{impurity}^2)\vert_{\epsilon^2=0}&=&-\sqrt{2}\epsilon^1\pa_t (\sigma \psi_2)\label{sec:intro:cur}\\
   \delta ( \mathcal{L}_{impurity}^1- \mathcal{L}_{impurity}^2)\vert_{\epsilon^1=0}&=&-\sqrt{2}\epsilon^2\pa_t (\sigma \psi_1)\label{sec:intro:cur2}
 \eea
 The total Lagrangian $\mathcal{L}_{tot}=\mathcal{L}_{0}-\mathcal{L}_{impurity}^1+ \mathcal{L}_{impurity}^2$ has the following form in components
 \be
 \mathcal{L}_{tot}\vert=\frac{1}{2}\pa_\mu \phi\pa^\mu\phi+\frac{1}{2}F^2+W_\phi(\phi)F-\sqrt{2}\phi_x \sigma+\sqrt{2}\sigma F.
 \ee
  After eliminating $F$ ($F=-W_\phi(\phi)-\sqrt{2}\sigma$) and for $\phi_t=0$ the on-shell Lagrangian reads
 \be
  \mathcal{L}_{tot,on-shell}\vert=-\frac{1}{2}\phi_x^2-\frac{1}{2}W_\phi^2(\phi)-\sqrt{2}W_\phi(\phi)\sigma-\sigma^2-\sqrt{2}\phi_x\sigma \label{sec:intro:fullL}
 \ee
 We recall that (\ref{sec:intro:fullL}) is invariant under one-half of the SUSY transformations (for $\epsilon^1=0$). As usual, the BPS equation can be obtained from (\ref{sec:intro:tpsi})
 \bea
 \delta\psi_1&=&i\epsilon^2(F-\phi_x)\label{sec:intro:varpsi1},\\
  \delta\psi_2&=&i\epsilon^1(-F-\phi_x)\label{sec:intro:varpsi2}.
 \eea
 Since in order to preserve one-half of SUSY at the level of the Lagrangian we have to impose $\epsilon^1=0$, the condition $ \delta\psi_2=0$ is automatically satisfied. From (\ref{sec:intro:varpsi1}) we obtain the condition
 \be
 F-\phi_x=0, \,\,\,\text{or}\,\,\, \phi_x=-W_\phi(\phi)-\sqrt{2}\sigma \label{sec:intro:BPS}.
 \ee
 It is easy to check that (\ref{sec:intro:BPS}) implies the Euler-Lagrange equations for (\ref{sec:intro:fullL}). Also, under the replacement
 \be
 W_\phi(\phi)\rightarrow \sqrt{2 U}
 \ee
the Lagrangian (\ref{sec:intro:fullL}) corresponds to (\ref{sec:intro:imp}).

The specific form of introducing the impurity preserving the BPS property gives us some hints on how to generalize this result to other models. Since the impurity breaks the translation invariance, one should expect that, if the model preserves part of the SUSY, the superalgebra contains only time translations (see Sec. \ref{central}). Now let us assume that the original model is BPS. From the SUSY point of view, this implies that the fermionic transformations (after a proper reduction of the parameter space) are either time derivatives or proportional to the BPS equations. In the previous case we had
\bea
\delta_\epsilon\psi_1&=&i\epsilon^2(F-\phi_x), \label{sec:intro:fert1}\\
\delta_\epsilon\psi_2&=&-i\epsilon^2\phi_t.\label{sec:intro:fert2}
\eea
On the other hand we have, in general 
\be
[\delta_\epsilon,\delta_\eta]X=-2i \epsilon^\alpha\eta^\beta\pa_{\alpha\beta}X.
\ee
Taking into account (\ref{sec:intro:fert1}) we have
\be
[\delta_\epsilon,\delta_\eta]\psi_1=2\delta_\epsilon\delta_\eta\psi_1=2i \eta^2 \delta_\epsilon(F-\phi_x)=-2i\eta^2\epsilon^2\psi_{1,t}
\ee
Therefore, after the reduction to the BPS space ($\epsilon^1=0$), the transformation of the off-shell BPS equation is a time derivative. This property can be stated as follows: the addition of a term of the form $\sigma(x)\Sigma$ (where $\Sigma=0$ is the off-shell BPS equation) to a SUSY BPS Lagrangian preserves the amount of supersymmetry preserved by the BPS solutions of the original model and, as a consequence, the BPS property.

\subsection{The central charge}
\label{central}

Let us consider again the model (\ref{sec:intro:imp}). As we discussed above, this model breaks explicitly one real supersymmetric generator. The supercharge is defined as follows
\be
Q_\alpha=\int dx\, J_\alpha^0,
\ee
where $J_\alpha^\mu$ is the supercurrent. The model (\ref{sec:intro:imp}) is $1/2$-supersymmetric provided that $\epsilon^1=0$ and this implies that $J_1^0=0\rightarrow Q_1=0$. Thus we have
\bea
&&\lbrace Q_2,Q_2\rbrace=2 P_0-2Z,\label{sec:central:alge}\\
&&\lbrace Q_1,Q_1\rbrace=0,\\
&&\lbrace Q_1,Q_2\rbrace=\lbrace Q_2,Q_1\rbrace=0.
\eea
Now, the impurity only adds an extra term proportional to the supercurrent (\ref{sec:intro:cur2}). A direct computation shows that
\be
Q_2=\int dx\left(\pa_t\phi\,\psi_1+\pa_x\phi\,\psi_2+\frac{1}{2}(F+W_\phi+\sqrt{2}\sigma)\,\psi_2\right).\label{sec:central:charge}
\ee
With the (anti)commutation relations
\be
[p(x),\phi(y)]=i\delta(x-y),\,\,\, \lbrace \psi_\alpha(x),\psi_\beta(y)\rbrace=\delta_{\alpha\beta}\delta(x-y).
\ee
Explicitly
\bea
P_0&=&\int dx\left(\frac{1}{2}\phi_t^2+\frac{1}{2}\phi_x^2+i\psi_{2,x}\psi_1+\frac{1}{2}W_\phi^2+\sqrt{2}W_\phi\sigma+\sigma^2+\sqrt{2}\phi_x \sigma\right)\\
Z&=&\int dx \phi_x W_\phi=W(\phi)\vert_{x=\infty}-W(\phi)\vert_{x=-\infty}
\eea
Since
\be 
\lbrace Q_2,Q_2\rbrace\geq 0,
\ee
we have from (\ref{sec:central:alge}) that the energy of a state $\vert \phi \rangle$ verifies the following inequality
\be
P_0 \geq Z.
\ee
This relation is clearly saturated for $Q_2\vert \phi \rangle=0$. For static solutions, this condition is equivalent to the equation (\ref{sec:central:charge})
\be
\phi_x-F=0\rightarrow \phi_x=-W_\phi-\sqrt{2}\sigma,
\ee
which is the BPS equation previously obtained. Note that if in the original model we replace $\sigma\rightarrow-\sigma$ the SUSY algebra becomes
\bea
&&\lbrace Q_2,Q_2\rbrace=0,\label{sec:central:alge1}\\
&&\lbrace Q_1,Q_1\rbrace=2 P_0-2Z,\\
&&\lbrace Q_1,Q_2\rbrace=\lbrace Q_2,Q_1\rbrace=0.
\eea
After a convenient change $W_\phi \rightarrow -W_\phi$ we get the saturating condition $\phi_x= W_\phi+\sqrt{2}\sigma$.


\subsection{The zero modes}
\subsubsection{The fermionic zero modes}
\label{ferm}

Let us consider the following model
 \be
 \mathcal{L}_{tot}\vert=\frac{1}{2}\pa_\mu \phi\pa^\mu\phi+\frac{i}{2}\psi^\alpha\pa_\alpha^{\,\,\beta}\psi_\beta+\frac{1}{2}F^2+W_\phi(\phi)F-\sqrt{2}\phi_x \sigma+\sqrt{2}\sigma F+\frac{1}{2}W_{\phi \phi} (\phi)\psi^\alpha\psi_\alpha.\label{sec:ferm:full}
 \ee
The fermion zero mode equations are given by
 \bea
\partial_x \psi_2-W_{\phi \phi}(\phi)\psi_2&=&0,\label{sec:ferm:fer1}\\
\partial_x  \psi_1+W_{\phi \phi}(\phi)\psi_1&=&0.\label{sec:ferm:fer2}
 \eea
In the standard case (without impurity), the fermionic zero mode is simply the derivative of the solitonic solution which obviously represents the translational zero mode of the soliton. Hence, the only normalizable solution for the soliton reads
\be
\psi_\alpha=\left(
\begin{matrix}
\phi^s_x(x)\\
0
\end{matrix}\right),
\ee
while for the antisoliton 
\be
\psi_\alpha=\left(
\begin{matrix}
0\\
\phi^a_x(x)
\end{matrix}\right).
\ee
In the presence of the impurity, the fermion zero mode equations are still given by (\ref{sec:ferm:fer1}) and (\ref{sec:ferm:fer2}) but there is only one (modified) BPS equation $\phi_x=-W_{\phi}-\sqrt{2}\sigma$. This leads to the following fermionic zero mode
\be
\psi_\alpha=\left(
\begin{matrix}
\phi_x (x)\exp\left(\sqrt{2} \int_0^xdx\frac{\sigma_x(x)}{\phi_x(x)}\right)\\
0
\end{matrix}\right).
\ee

\subsubsection{The bosonic zero modes}
\label{bos}

The linear fluctuation equation in the kink/antikink background can be derived by inserting the decomposition $\phi(t,x)=\phi_c(x)+\cos(\omega t)\eta(x)$, where $\phi_c$ is a kink/antikink solution. The resulting fluctuation equation is
\be
-\eta_{xx}(x)+\left(W_{\phi }(\phi_c)W_{\phi \phi \phi}(\phi_c)+W_{\phi \phi}(\phi_c)^2+\sqrt{2}W_{\phi \phi \phi}(\phi_c)\sigma\right)\eta(x)=\omega^2\eta(x).
\ee
We will show that the bosonic and fermionic zero modes coincide. Let us assume that $\eta$ satisfies the fermionic zero mode equation (\ref{sec:ferm:fer2})
\be
\eta_x=-W_{\phi \phi} \eta
\ee
where $\phi$ is a BPS solution i.e., obeys
\be
\phi_x=-W_\phi-\sqrt{2}\sigma
\ee
Then, acting with $\partial_x$ we get
\be
\eta_{xx}=-W_{\phi \phi \phi} \phi_x \eta - W_{\phi \phi} \eta_x= W_{\phi \phi \phi} (W_\phi+\sqrt{2}\sigma) \eta + W_{\phi \phi} W_{\phi \phi} \eta
\ee
where we used the Bogomolnyi equation and the fermionic zero mode equation. But the last formula is exactly the bosonic mode equation. Hence, as one could expect, both modes exactly coincide. 

\vspace*{0.2cm}

Here we present an example of the BPS-impurity model which allows for the exact computation of the BPS soliton as well as the related zero mode. Let us take $W_\phi=(1-\phi^2)$ which corresponds to the $\phi^4$ potential while the impurity is
\be
\sigma (x)= \frac{\alpha-1}{\sqrt{2}} \frac{1}{\cosh^2 \alpha x}
\ee
where $\alpha$ is a real parameter, $\alpha \neq 0$. Then, the Bogomolnyi equation $\phi_x=-W_\phi -\sqrt{2}\sigma$ has the following exact solution
\be
\phi = -\tanh \alpha x
\ee
which is a kink (positive topological charge) for $\alpha < 0$. For $\alpha >0$ we get an antikink (negative topological charge). For $\alpha =1$ we arrive at the usual $\phi^4$ theory antikink.   
The zero mode can be also explicitly found and reads
\be
\eta = - \frac{\alpha}{\left(\cosh \alpha x \right)^{2/\alpha}}.
\ee

\subsection{Generalizing the BPS preserving impurity}
\label{gen}

As a matter of fact, the addition of the impurity preserving half of the BPS structure is not unique. As an example, let us assume the following, new impurity term 
\be
\mathcal{L}_{impurity}=-\frac{1}{2}\sigma \phi_x^2 \label{sec:exam:nont}
\ee
which is added to the standard bosonic, impurity free part of the model. As we will see, it is again possible to include some extra terms that will preserve a part of the BPS structure. The form of such new terms can be deduced using our supersymmetric approach. From the supersymmetric point of view, the impurity term in (\ref{sec:exam:nont}) suggests a term of the form
\be
\mathcal{L}_1=\frac{1}{4}\int d^2\theta \sigma D^\alpha \Phi D_\alpha \Phi.
\ee
But because of the reasons discussed above, this term does not preserve supersymmetry, i.e., 
\be
\delta\mathcal{L}_1\vert_{\epsilon^2=0}=\sigma\pa_x\left(\epsilon^1\psi_1(\phi_x+F)-\epsilon^1\psi_2 \phi_t\right)\label{sec:gen:var},
\ee
which is not a total derivative. Note that the time component of (\ref{sec:gen:var}) is a total derivative since $\pa_t \sigma(x)=0$. On the other hand, we introduce
\be
\mathcal{L}_2=\sigma_x\phi(F+\phi_x),
\ee
with the following transformation properties
\be
\delta\mathcal{L}_2\vert_{\epsilon^2=0}=-\sigma_x \epsilon^1\psi_1(F+\phi_x)-\sigma_x\phi \epsilon^1\psi_{2,t}.\label{sec:gen:var2}
\ee
Again, if we combine (\ref{sec:gen:var}) and (\ref{sec:gen:var2}) we have the following term
\be
\delta(\mathcal{L}_1-\mathcal{L}_2)\vert_{\epsilon^2=0}= \epsilon^1\pa_x\left( \sigma \psi_1(F+\phi_x)\right)+\epsilon^1\pa_t (\sigma_x\phi\psi_2).
\ee
As a consequence, ($\mathcal{L}_1-\mathcal{L}_2$) preserves half of the supersymmetry. After including a linear $\sigma$-model term and a superpotential, the full Lagrangian (for $\phi_t=0$) reads
\be
\mathcal{L}_{tot}\vert=\mathcal{L}_{0}+\mathcal{L}_1-\mathcal{L}_2\vert=-\frac{1}{2}\phi_x^2\left(1+\sigma\right)+\frac{1}{2}F^2(1+\sigma)+\sigma_x\phi(F+\phi_x)+F W'(\phi).
\ee
Solving for $F$
\be
F=\frac{\sigma_x \phi-W'(\phi)}{1+\sigma},\label{sec:exam:aux}
\ee
the on-shell Lagrangian takes the form
\be
\mathcal{L}_{tot, \; on-shell}=-\frac{1}{2}\phi_x^2(1+\sigma)-\sigma_x\phi_x\phi-\frac{1}{2}\frac{\left(W'(\phi)-\sigma_x\phi\right)^2}{1+\sigma}.\label{sec:exam:full}
\ee
Finally, taking into account (\ref{sec:intro:tpsi}) and (\ref{sec:exam:aux}), the BPS equation can be expressed as
\be
\phi_x+\frac{\sigma_x \phi-W'(\phi)}{1+\sigma}=0.
\ee

It is clear from (\ref{sec:exam:full}) that the addition of an impurity (in the last example simply, $\sigma\phi_x^2$) and the conservation of part of the BPS structure can lead to nontrivial  Lagrangians.  As we have shown here, the latter condition can be translated into a partial  explicit breaking of the SUSY generators of an underlying supersymmetric model.  As a consequence of this reasoning, one can construct a large family of (half) BPS preserving soliton-impurity models which in the limit $\sigma \rightarrow 0$ reproduce the pure (no impurity) model
\be
\mathcal{L}_0=\frac{1}{2}( \partial_\mu \phi)^2 - U.
\ee
A general coupling to an impurity $\sigma$ preserving half of the BPS-ness has the following form
\be
\mathcal{L}_\sigma=  \frac{1}{2}H^2( \partial_\mu \phi)^2 - \frac{U}{H^2}  - G^2 - 2 \frac{\sqrt{U}G}{H} - \sqrt{2}HG \phi_x
\ee
where $H, G$ are functions of $\phi$ and $\sigma$ such that for the vanishing impurity i.e., when $\sigma \rightarrow 0$, $H(\phi, \sigma) \rightarrow 1$ and $G(\phi,\sigma)\rightarrow 0$ to recover the pure bosonic model. The static energy reads
\bea
E&=&\int dx \left[  \frac{1}{2}H^2 \phi_x^2 +\frac{U}{H^2}  + G^2 + 2 \frac{\sqrt{U}G}{H} + \sqrt{2}HG\phi_x \right] \\
&=& \int_{-\infty}^\infty dx \left( \frac{H}{\sqrt{2}} \phi_x + \frac{\sqrt{U}}{H} +G \right)^2 - \sqrt{2} \int_{-\infty}^\infty dx \phi_x \sqrt{U} \\
&\geq&  - \sqrt{2} \int_{-\infty}^\infty dx \phi_x \sqrt{U} = - \sqrt{2} \int_{\phi(-\infty)}^{\phi(\infty)} d\phi \sqrt{U} =  - Q \sqrt{2} \int_{\phi_+^v}^{\phi_-^v} d\phi \sqrt{U}
\eea
The bound is saturated if the Bogomolnyi equation is obeyed
\be
\frac{H}{\sqrt{2}} \phi_x + \frac{\sqrt{U}}{H} +G =0.
\ee
For $H=1$ and $G=\sigma$ we get the first BPS-impurity model, while $H=\sqrt{1+\sigma}$, $G=\frac{\sigma_x}{\sqrt{1+\sigma}}$ leads to the second model (up to a factor of $\sqrt{2}$). Let us remark that such a generalisation reminds us of the construction of generalised BPS models as presented in \cite{k-model-2}, \cite{adam-santamaria}.

Note, that $G$ is a completely arbitrary function of the field as well as the impurity (and can depend, for example, on its higher derivatives). As we will see in the next section, this result can be understood in a more general framework.

\section{ $\mathcal{N}=1$ SUSY BPS impurities - a general construction } \label{sect-3}
The standard $\mathcal{N}=1$ SUSY algebra in $1+1$ dimensions has the following form
\be
[\delta_\epsilon,\delta_\eta]X=-2i \epsilon^\alpha\eta^\beta\pa_{\alpha\beta}X.
\ee
After the restriction to the BPS subspace, $\epsilon^2=0$ (or $\epsilon^1=0$) it reduces to 
\be
[\delta_\epsilon,\delta_\eta]X=-2i \epsilon^1\eta^1\pa_t X.
\ee
Since the BPS subspace has only one SUSY parameter we have the following property
\be
[\delta_\epsilon,\delta_\eta]X=2\delta_\epsilon\delta_\eta X=-4 i\epsilon^1\eta^1\pa_t X.
\ee
As a consequence, for $Y=\delta X$, i.e. $Y\in\text{Im}(\delta)$, we have: if $Y\in \text{Im}(\delta)\Rightarrow \delta_\eta Y\propto\pa_t X$, or in words, the SUSY transformation of a term in the image of the SUSY transformation gives a time derivative. This has a nice consequence for the models with impurities, since the addition of any term of the form $\sigma (x) Y$ with $Y\in \text{Im}(\delta)$ will preserve the amount of SUSY preserved by the BPS soliton of the original model. Therefore, if $\mathcal{L}$ is a BPS model with $\mathcal{N}=1$ SUSY, then $\forall X$, the following Lagrangian is also BPS
\be
\tilde{\mathcal{L}}=\mathcal{L}+\sigma(x)\delta_\epsilon X\vert 
\ee
where $\vert$ means that the SUSY parameter has been removed. The following question arises naturally: is it possible to add a term $Y$, $Y\not\in \text{Im}(\delta)$ and such that 1/2 of SUSY is preserved? Let us assume that the statement is true. In order to have $1/2$ SUSY the following must hold
\be
\delta\tilde{\mathcal{L}}=\delta\mathcal{L}+\sigma(x)\delta Y=\epsilon^1 \pa_\mu j^\mu_1+\sigma(x)\epsilon^1\pa_t J.
\ee
But if we choose $\chi=\int dt \delta Y=\epsilon^1 J$, we have $\delta\chi=\int dt \delta\delta Y\propto Y$, and therefore $Y\in \text{Im}(\delta)$. On the other hand, if $\delta Y=0$ then $Y\not\in \text{Im}\delta$. But if $\delta Y=0$, Y does not depend on the fields and, therefore, Y must be a constant (or an impurity). Let us take $Y=\mu(x)$, then if $\sigma(x)\mu(x)\in L^1(\mathbb{R})$, the model $\tilde{\mathcal{L}}$ is still BPS, but the topological bound is shifted by a amount $\Delta=\int dx \sigma(x)\mu(x)$ (the effect of the impurity is trivial: it does not affect the e.o.m.). 

Nontrivial situation: $Y\in \text{Im}(\delta)$. Since $Y$ contains bosonic terms (non-vanishing when $\psi_\alpha\rightarrow 0$), it has to be the image of a term linear in fermions ($X$). We have two possibilities
\be
X=\psi_1 H(F,\phi,\mu(x)), \,\, \text{or} \,\, X=\psi_2 H(F,\phi,\mu(x)),
\ee
where $\mu(x)$ is another impurity. From the SUSY variation of fermions in the restricted subspace, we obtain
\bea
Y&=&\delta \psi_1 H(F,\phi,\mu(x))+(\text{fermions})\propto \pa_t\phi H(F,\phi,\mu(x))+(\text{fermions}) \label{gen1}\\
&&\text{or}\nonumber\\
Y&=&\delta \psi_2 H(F,\phi,\mu(x))+(\text{fermions})\propto (F+\pa_x\phi) H(F,\phi,\mu(x))+(\text{fermions})\label{gen2}.
\eea

In both cases the topological bound does not change: for (\ref{gen1}), because the static energy functional does not change, while in the second, because the modification is proportional to the BPS equation. Note also that, depending on the form of $H$, it may be necessary to add fermionic terms in order to preserve $1/2$ SUSY. We can gather these results in the following lemma: 

\begin{lemma}
Let $\mathcal{L}$ be a $1+1$ dimensional scalar model with $\mathcal{N}=1$ SUSY and a topological bound $T$. Let $\tilde{\mathcal{L}}$ be a BPS preserving impurity model based on $\mathcal{L}$ of the form $\tilde{\mathcal{L}}=\mathcal{L}-\sigma(x)Y$ with topological bound $\tilde{T}$. Then all possible impurity terms $Y$, preserving one supersymmetric generator (and one BPS sector) are of one of the following forms:\\
i) $Y\in\text{Im}(\delta)\,(=\delta X, \forall X)$ and $\tilde{T}=T$.\\
ii)$Y\not\in\text{Im}(\delta)\,(=\mu(x))$ and $\tilde{T}=T+\Delta$, with $\Delta=\int dx \,\sigma(x)\mu(x)$.
\end{lemma}


\section{The impurity baby BPS Skyrme model}
Having systematically investigated the BPS preserving impurities in (1+1) dimensions, we want to apply the same supersymmetric approach to implement such impurities also in (2+1) dimensions. 
\subsection{The model}

We begin our analysis of (2+1) dimensional theories with the baby BPS Skyrme model \cite{bBPS}, 
which provides also a lower dimensional counterpart of the BPS Skyrme model \cite{BPS Sk}. The static energy reads
\be
E_{baby BPS}=\int d^2 x \left[ \lambda^2 B_0^2 + \mu^2 \mathcal{U} \right]
\ee
where $\mathcal{B}_0$ is a temporal component (topological charge density) of the pertinent topological current 
\be
\mathcal{B}^\mu=\frac{1}{8\pi}  \epsilon^{\mu \nu \rho} \vec{\phi} \cdot \left( \partial_\nu \vec{\phi} \times \partial_\rho \vec{\phi} \right)
\ee
such that $B=\int d^2 x \mathcal{B}_0$ is the topological charge. 
Furthermore, $\vec{\phi} \in \mathbb{S}^2$ is a three component isovector with the unit length $\vec{\phi}^2=1$ and $\mathcal{U}$ is a non derivative part, that is, a potential (with at least one isolated vacuum). It is known that the model possesses the following topological bound (where $d\Omega_{\mathbb{S}^2}$ is the "volume" (area) form on the (target space) unit sphere)
\be
E_{baby BPS} \geq 2\mu \lambda |B|  \left\langle \sqrt{\mathcal{U}} \right\rangle, \;\;\;  \left\langle \sqrt{\mathcal{U}} \right\rangle \equiv \frac{1}{\mbox{vol}_{\mathbb{S}^2}} \int d\Omega_{\mathbb{S}^2} \sqrt{\mathcal{U}}.
\ee
The bound is saturated if the following Bogomolnyi equation is obeyed,
\be
\lambda \mathcal{B}_0  \pm \mu \sqrt{\mathcal{U}} =0.
\ee
Solutions of this equation exist in any topological sector and can be found in an exact form once a potential is chosen. This model is a higher dimensional counterpart of the standard scalar model in (1+1) dimensions. Indeed, both Lagrangians are sums of a potential (a function of the fields) and the square of a topological current. 

Again, it is possible to couple an impurity in such a way that one-half of the BPS property of the model remains preserved. Namely, the energy functional reads 
\be
E=\int d^2 x \left[ \lambda^2  \mathcal{B}_0^2 + \mu^2 \mathcal{U} - 2 \lambda \sigma \mathcal{B}_0 +2\mu \sigma \sqrt{\mathcal{U}} + \sigma^2\right]. \label{sec:baby:ener}
\ee
Then we find the topological bound
\bea
E&=&\int d^2 x \left( \lambda \mathcal{B}_0 - \mu \sqrt{\mathcal{U}}-\sigma \right)^2 + 2 \lambda \mu  \int d^2 x  \mathcal{B}_0 \sqrt{\mathcal{U}} \\
&\geq& 2 \lambda \mu  \int d^2 x \mathcal{B}_0 \sqrt{\mathcal{U}} =  2 \lambda \mu  B \left\langle \sqrt{\mathcal{U}} \right\rangle
\eea
which is saturated if the following modified, impurity dependent, Bogomolnyi equation is obeyed,
\be
\lambda  \mathcal{B}_0 - \mu \sqrt{\mathcal{U}} - \sigma=0.
\ee
One can easily prove that the Bogomolnyi equation implies the full Euler-Lagrange equation, see appendix \ref{a1}. 

As an example, we consider the potential $\mathcal{U}=(1-\phi^3)^2$. Then, we can find an exact solution of the Bogomolnyi equation for any radial impurity $\sigma = \sigma(r)$. In fact, introducing, via the usual stereographic projection, a complex field $u$
\be
u=\frac{\phi^1-i\phi^2}{1+\phi^3}
\ee
and assuming $u=\sqrt{\frac{g}{1-g}} e^{-in\varphi}$, where $r,\varphi$ are the polar coordinates and $n$ is a positive integer, the Bogomolnyi equation takes a linear form
\be
-\lambda \frac{n}{2\pi} g_y=2\mu g + \sigma
\ee
Here, for convenience we use $y=r^2/2$. The topologically nontrivial boundary conditions are $g(0)=1$, $g(\infty)=0$. Then, for an exponentially localized impurity $\sigma = \alpha e^{-\beta y}$ we find
\be
g(y)= e^{-\frac{4\pi \mu }{n \lambda} y} + \frac{2\pi}{n\lambda}  \frac{\alpha}{\beta -\frac{4\pi \mu}{n \lambda}} \left( e^{-\beta y} - e^{-\frac{4\pi \mu }{n \lambda} y}\right)
\ee
Observe that  for $\beta = \frac{4\pi \mu }{n \lambda}$, which might occur only for one value of the topological charge, the solution takes a different form, 
\be
g(y)=\left(1+\frac{2\pi\alpha}{\lambda n} y \right) e^{-\frac{4\pi \mu }{n \lambda} y}.
\ee
The corresponding topological charge is positive, $B=n>0$. Anti baby Skyrmions, i.e., solitons with negative topological charge, do not obey the Bogomolnyi equation and can only be found after solving the full Euler-Lagrange equations. This is, of course, a much more difficult task. 


\subsection{SUSY in the impurity baby BPS Skyrme model}

We can use the same strategy as in Sec. \ref{BPS} to build a SUSY BPS baby Skyrme model preserving the BPS structure. This model has a natural $\mathcal{N}=2$ SUSY formulation \cite{k-susy-8}, which can be used to introduce the impurity term. The BPS solitons in the original BPS baby Skyrme model satisfy the following BPS condition
\be
\delta\psi=0,\,\,\, \delta\bar{\psi}=0.\label{sec:baby:fert}
\ee

The component expansion of the SUSY BPS baby Skyrme model (after stereographic projection) has the form
\be
\mathcal{L}_{\text{baby}}\vert=g(u,\bar{u})\left(\pa_\mu\bar{u}\pa^\mu u+F\bar{F}\right)+h(u,\bar{u})\left((\pa_\mu u)^2(\pa_\nu\bar{u})^2+2F\bar{F}\pa_\mu \bar{u}\pa^\mu u+(F\bar{F})^2\right),
\ee
where $h(u\bar{u})=1/(1+\vert u\vert^2)^4$. In order to reproduce (\ref{sec:baby:ener}) we introduce
\be
\mathcal{L}_{\text{impurity}}=2\sigma(x)\sqrt{h(u,\bar{u})}\left(-\pa_i u\pa_i\bar{u}+F\bar{F}+i(\pa_1\bar{u}\pa_2 u-\pa_1 u\pa_2\bar{u})\right).\label{sec:baby:impu}
\ee
Note that the last term is the topological charge density. In terms of the (anti)holomorphic derivative, (\ref{sec:baby:impu}) reads
\be
\mathcal{L}_{\text{impurity}}=2\sigma(x)\sqrt{h(u,\bar{u})}\left(F\bar{F}-\pa_z u \pa_{\bar{z}}\bar{u}\right).\label{sec:baby:impuh}
\ee

The addition of (\ref{sec:baby:impu}) will preserve part of the SUSY if it belongs to the image of the restricted SUSY transformation (see Sec. \ref{s:N2}). It is possible to verify that the following term is the preimage of (\ref{sec:baby:impuh}) modulo fermionic terms
\bea
X&=& \sqrt{h(u,\bar{u})}\left(\psi_{\bar{z}}\left(\pa_{\bar{z}}\bar{u}+\bar{F}e^{i\gamma}\right)+\bar{\psi}_z\left(\pa_z u+Fe^{-i\gamma}\right)\right),\\
\delta X\vert&=&2\sqrt{2}\,\bar{\epsilon}_{\dot{2}}\,\mathcal{L}_{\text{impurity}}+(\text{fermions}).
\eea

We have finally
\bea
\mathcal{L}_{\text{baby}}+\mathcal{L}_{\text{impurity}}\vert&=&h(u,\bar{u})\left((\pa_\mu u)^2(\pa_\nu\bar{u})^2 -(\pa_\mu\bar{u}\pa^\mu u)^2\right)-\frac{g^2(u,\bar{u})}{4 h(u,\bar{u})}-\sigma(x)\frac{g(u,\bar{u})}{\sqrt{h(u,\bar{u})}}+\nonumber\\
&+&2\sigma(x)\mathcal{B}_0\sqrt{h(u,\bar{u})}-\sigma^2(x)
\eea
which corresponds to (\ref{sec:baby:ener}) after the identification $\frac{g^2(u,\bar{u})}{4 h(u,\bar{u})}=\mathcal{U}$ and $\lambda=\mu=1$.

\section{The impurity $CP^1$ model}
\label{sec-cp1}
\subsection{The model}

Now we can turn to the simplest solitonic model in (2+1) dimensions (which is also integrable in its static version), that is, the $CP^1$ model
\be
\mathcal{L}_{CP^1}=\frac{1}{2} \frac{ \partial_\mu u \partial^\mu \bar{u}}{(1+|u|^2)^2}
\ee
whose static energy reads
\be
E_{CP^1}= \int d^2 z \frac{1}{(1+|u|^2)^2} \left( u_z \bar{u}_{\bar{z}} + u_{\bar{z}} \bar{u}_z \right)
\ee
where we use the complex plane coordinate $z=x+iy$. 
Topologically nontrivial solutions are just (anti)holomorphic maps of degree $n$, which obey
the Bogomolnyi equation (here the Cauchy-Riemann (CR) equations)
\be
u_z=0 \;\;\; \mbox{or} \;\;\; u_{\bar{z}}=0.
\ee
For solutions of the CR equations, the energy equals the modulus of the topological charge (the degree of the map)
\be
Q=\frac{1}{\pi} \int d^2 z \frac{1}{(1+|u|^2)^2} \left( u_z \bar{u}_{\bar{z}} - u_{\bar{z}} \bar{u}_z \right)
\ee

The BPS preserving $CP^1$ model with an impurity is defined as follows (we restrict ourselves only to the static energy functional)
\be
E=E_{CP^1} + E_{impurity} \label{cp-imp}
\ee
where
\be
E_{impurity}= \int d^2 z \frac{1}{(1+|u|^2)^2} \left( 2 \sigma \bar{\sigma} +2 \sigma u_z +2 \bar{\sigma} \bar{u}_{\bar{z}}  \right) 
\ee
Then, 
\bea
E&=&\int d^2 z \frac{1}{(1+|u|^2)^2} \left[ 2 \left( u_z+\bar{\sigma} \right)\left( \bar{u}_{\bar{z}} +\sigma\right)  - u_z \bar{u}_{\bar{z}} + u_{\bar{z}} \bar{u}_z\right] \\
&\geq& \int d^2 z \frac{1}{(1+|u|^2)^2} \left[  - u_z \bar{u}_{\bar{z}} + u_{\bar{z}} \bar{u}_z\right] = -\pi Q .\\ 
\eea
The topological bound is saturated if the following Bogomolnyi equation is satisfied,
\be
u_z+\bar{\sigma}=0.
\ee
This is an inhomogeneous generalization of the anti-holomorphic CR equation. One can check that solutions of this equation obey the full EL equation, i.e., are the critical points of (\ref{cp-imp}). As the Bogomolnyi equation is a non-homogeneous linear differential equation the BPS sector is still {\it completely solvable}. This reminds us of the situation of the so-called integrable defect, that is, an impurity which does not affect the integrability of the underlying scalar field theory (like sine-Gordon) \cite{integrability}. Here, a solitonic-impurity solution consists of the homogeneous part, which is still given by an arbitrary antiholomorphic map $f(\bar z)$ and a pertinent unique solution of the non-homogenous part. Therefore, contrary to the pure $CP^1$ model, the total solution does not have to be purely antiholomorphic (or holomorphic). Of course, this strongly depends  on the impurity. Furthermore, the impurity may also contribute to the topological charge (degree) $Q$ of the full soliton-impurity solution.

Let us consider an antiholomorphic impurity, for example, $\sigma_m=B\bar{z}^m$, where $m\in \mathbb{N}$ and $B$ is a complex constant. Hence, a BPS solution is 
\be
u= A\frac{(\bar{z}-\bar{z}_1)...(\bar{z}-\bar{z}_r)}{(\bar{z}-\bar{z}_1)...(\bar{z}-\bar{z}_s)} - \frac{\bar{B}}{m+1} z^{m+1}
\ee
where we assumed a rational map as a solution of the homogenous part of the Bogomolnyi equation. Here the polynomials in the numerator and denominator do not have common divisors. Furthermore, max$(r,s)=n$ and $A\in \mathbb{C}$. However, the computation of the degree of the solution is not a completely trivial task. Even in the case of the constant impurity $m=0$ and the linear antiholomorphic part the topological degree of the solution 
\be
u(z,\bar{z})=A\bar{z} -\bar{B}z
\ee
depends on the constants $A, B$ \cite{Sch-DM}. Concretely
\be
\mbox{deg } u = \left\{
\begin{array}{rr}
1 \;\;\;\; & |A| < |B| \\
0 \;\; \;\; & \mbox{if} \;\;\;\; |A|=|B| \\
-1 \;\;\;\;& |A|>|B|
\end{array}
\right.
\ee

This observation can have a nontrivial impact on the moduli space of soliton-impurity solutions, as different values of the parameters of the homogeneous part of a solution (with a fixed degree) can lead to different degrees of the total solution and, therefore, to energetically inequivalent configurations. We will address this problem (and its dependence on a particular choice of the impurity) in a separate paper. 

We also remark that our constant impurity leads to a Bogomolnyi equation which is identical to the Bogomolnyi equation very recently found for the magnetic planar Skyrmions with the Dzyaloshinskii-Moriya interaction energy \cite{Sch-DM}. This may suggest that there is a deeper relation between our impurity coupling and Dzyaloshinskii-Moriya like terms \cite{DM}.  Of course, this is of high importance as far as a possible experimental realization of the BPS preserving impurity is concerned. Moreover, one can ask whether  there exist DM counterparts of our non-homogenous CR Bogomolnyi equations also in the case of more complicated (non-constant) BPS preserving impurities. 

\vspace*{0.2cm}

A related issue is the global $U(1)$ symmetry. Of course, for an impurity which trivially transforms under this group, the model (\ref{cp-imp}) is no longer invariant under the global $U(1)$. However, this symmetry is effectively restored in the BPS sector. Indeed, the group can act nontrivially on the homogeneous part of BPS solutions leading to energetically equivalent configurations. 

A possibility to restore the global $U(1)$ symmetry completely is to assume that the impurity also transforms in the fundamental representation of it i.e., $\sigma \rightarrow e^{i\varphi} \sigma$, where $\varphi$ is the transformation parameter. This may apply to impurities which originate as a spatially frozen lump, which is a mechanism proposed in \cite{vortices} for the magnetic impurity in the Abelian Higgs model. 

Finally, one can introduce an impurity term which is $U(1)$ invariant, although the impurity transforms trivially. For example one can consider the following modification 
\be
\tilde{E}_{impurity}= \int d^2 z \frac{1}{(1+|u|^2)^2} \left( 2 \sigma \bar{\sigma} u\bar{u}+2 \sigma \bar{u} u_z +2 \bar{\sigma} u \bar{u}_{\bar{z}}  \right) \label{cp-imp-inv}
\ee
The resulting Bogomolnyi equation is also deformed and reads 
\be
u_z+ u \bar{\sigma}=0.
\ee
However, one can still find exact solutions, where now the impurity term acts multiplicatively on the original antiholomorphic map
\be
u= A\frac{(\bar{z}-\bar{z}_1)...(\bar{z}-\bar{z}_r)}{(\bar{z}-\bar{z}_1)...(\bar{z}-\bar{z}_s)}  e^{-\int dz \bar{\sigma}}
\ee

A particular choice for the treatment of the global $U(1)$ transformation depends, of course, on the physical application one has in mind. Obviously, it will affect the possibility to promote this global symmetry to a local one, that is, the gauging of the impurity model.


\subsection{SUSY in the impurity $CP^1$ model}
\label{sub:CP}

The $\mathcal{N}=2$ formulation of the $CP^1$ model is well-known. It can be constructed in terms of the following K\"ahler potential
\be
\int d^2\theta d^2\bar{\theta}\,\text{ln}(1+\Phi^\dagger\Phi),
\ee
where $\Phi(\Phi^\dagger)$ are chiral (antichiral) superfields. Inspired by our one-dimensional construction is not very difficult to guess the SUSY form of (\ref{cp-imp}): 
\be
\mathcal{L}_{\text{impurity}}=\frac{1}{(1+\vert u \vert^2)^2}\sigma(x)(F-\pa_{\bar{z}} u)+\text{h.c.}+\text{fermions}.\label{sub:imp1}
\ee
In addition, this term lies in the image of $\delta$, as can be seen explicitly by taking the following transformation
\be
\mathcal{L}_{\text{impurity}}=\sigma(x)\,\delta\left(\frac{1}{(1+\vert u \vert^2)^2}\psi_1+\text{h.c.}\right)+\text{h.c.}\vert.
\ee
As we will see (Sec. \ref{s:N2}), this allows us to conclude that both the impurity (\ref{sub:imp1}) and the BPS solutions of the model preserve $1/4$ of SUSY.

\section{The impurity Abelian Higgs model at critical coupling}
\subsection{Magnetic impurity}
It was observed by Tong and Wong \cite{vortices} that the standard Abelian Higgs model at the critical
coupling admits a half-BPS preserving impurity extension. The impurity couples to the magnetic field $B$ and therefore is referred to as the magnetic impurity $\sigma_m$,
\be
E_{m}=\frac{1}{2} \int d^2 x \left( B^2 +\overline{D_i u} D_i u +\frac{1}{4}(1+\sigma_m-|u|^2)^2 -\sigma_m B \right)
\ee
Here, $F_{\mu \nu}=\partial_\mu A_\nu - \partial_\nu A_\mu$ is the field strength of the $U(1)$ gauge field $A_\mu$ and $D_\mu u=\partial_\mu u - iA_\mu u$ is the covariant derivative of the complex Higgs field $u$. The pertinent Bogomolnyi equations read
\bea
D_x u +iD_y u&=&0 \\
B -\frac{1}{2}(1- |u|^2)&=& \frac{1}{2}\sigma_m,
\eea
and its solutions saturate the following energy bound
\be
E_m \geq \pi N, \;\;\; N=\frac{1}{2\pi} \int d^2 x B.
\ee

\subsection{Higgs impurity}
As we have shown in Sec. \ref{sec-cp1}, the $CP^1$ model couples to an impurity in a half-BPS preserving manner, provided that the original CR equations get modified to an inhomogeneous version. The resulting model preserves the global U(1) symmetry (not only in the BPS sector) if the impurity is assumed to transform in the fundamental representation. This opens a new path to implement a partially BPS preserving impurity in the Abelian Higgs model. Now, contrary to the above case, the impurity multiplies the derivatives of the matter field and therefore we call it a Higgs impurity $\sigma_h$. Specifically, the model is
\be
E_{h}=E_{AH}+E_{impurity}
\ee
where to the standard Abelian Higgs part
\be
E_{AH}=\frac{1}{2} \int d^2 x \left( B^2 +\overline{D_i u} D_i u +\frac{1}{4}(1-|u|^2)^2 \right)
\ee
we add the impurity term
\be
E_{impurity}= \frac{1}{2} \int d^2 x \left( \sigma_h \bar{\sigma}_h + \sigma_h \overline{(D_x u + i D_y u)} + \bar{\sigma}_h (D_x u + i D_y u) \right)
\ee
which, after using complex coordinates $z, \bar{z}$, is a gauged version of the $CP^1$ impurity term
\be
E_{impurity}= \frac{1}{2} \int d^2 z \left(\frac{1}{2} \sigma_h \bar{\sigma}_h + \bar{\sigma}_h D_{\bar{z}} u + \sigma_h \overline{D_z u} \right) \label{higgs1}
\ee
The lower energy bound reads
\bea
E_h &=& \frac{1}{2} \int d^2 x \left( \left( B -\frac{1}{2}(1-u\bar{u}) \right)^2 +   \overline{(D_x u + i D_y u + \sigma_h)}(D_x u + i D_y u + \sigma_h)  \right) \\
&+& \frac{1}{2} \int d^2 x B -\frac{i}{2} \int d^2 x  \left( \partial_x (\bar{u} D_y u) - \partial_y (\bar{u} D_z u)\right) \\
&\geq & \pi N
\eea
where the last term in the second line integrates to zero. The bound is saturated if the following Bogomolnyi equations are satisfied
\bea
D_x u +iD_y u+\sigma_h&=&0 \\
B -\frac{1}{2}(1- |u|^2)&=&0.
\eea
As expected, this impurity enters only into the gauged CR equations, leaving the magnetic field equation unchanged. Of course, one should remember that this construction makes sense only if the local $U(1)$ transformation continues to be a symmetry also after the inclusion of the impurity. This implies that $\sigma_h \rightarrow e^{i\varphi (x,y)} \sigma_h$. As in the case of the $CP^1$ model, this can be relevant for an impurity originating from a spatially frozen vortex. 

It is also possible to implement a partially BPS preserving impurity term which is invariant under the gauge group even for a trivially transforming impurity, $\sigma_h \rightarrow  \sigma_h$. For example, one can use the globally $U(1)$ invariant $CP^1$ model with impurity as a hint. Hence,
\be
\tilde{E}_{impurity}= \frac{1}{2} \int d^2 z \left(\frac{1}{2} \sigma_h \bar{\sigma}_h u\bar{u}+ \bar{\sigma}_h \bar{u} D_{\bar{z}} u + \sigma_h u\overline{D_z u} \right) \label{higgs2}
\ee
This leads to the the following modification of the gauge CR equations,
\be
D_x u +iD_y u+ u \sigma_h=0.
\ee
We comment that, in contrast to the magnetic impurity, the Higgs impurity does not have any impact on the corresponding Bradlow law. 
%
%
Note that both impurities can be implemented in a completely independent way which results in the most general partially BPS preserving impurity Abelian Higgs model
\be
E_{m, \; h} = E_{magnetic} + E_{impurity}
\ee
This model possesses the same topological energy bound as the above impurity models, and the Bogomolnyi equation is of the following form,
\bea
D_x u +iD_y u+\sigma_h&=&0 \\
B -\frac{1}{2}(1+\sigma_m - |u|^2)&=&0.
\eea
The above remarks on the gauge invariance apply.


\subsection{SUSY in the impurity Abelian-Higgs model}

The magnetic impurity can be introduced in a $1/2$ SUSY invariant way as long as it is compensated by the auxiliary field of the vector multiplet \cite{vortices,mirror}. In its simplest form, it can be written as 
\be
\mathcal{L}^h_{\text{impurity}}=\sigma_m(x)(B-D).\label{sub:AH:im}
\ee
It is easy to see that (\ref{sub:AH:im}) corresponds to the image of $\lambda_\alpha$ (the fermion in the vector multiplet) under the gauge invariant SUSY transformation. It should be noted that (\ref{sub:AH:im}) is proportional to the gauged BPS equation and, therefore, the addition of the magnetic impurity cannot change the Bogomolnyi bound. The Higgs impurity, on the other hand, takes the form
\be
\mathcal{L}^h_{\text{impurity}}=\frac{1}{(1+\vert u \vert^2)^2}\sigma_h(x)(F-D_{\bar{z}} u)\bar{u}+\text{h.c.}+\text{fermions},\label{sub:AH:h}
\ee
which is an obvious generalization of (\ref{sub:imp1}), up to a factor. But, as in the impurity $CP^1$ model, the impurity (\ref{sub:AH:h}) only preserves $1/4$ of the SUSY. Since the subalgebra preserved by (\ref{sub:AH:h}) is contained in the subalgebra preserved by (\ref{sub:AH:im}), the presence of both impurities breaks the SUSY to $1/4$. It is interesting to note that (\ref{sub:AH:im}) and (\ref{sub:AH:h}) are perhaps the simplest nontrivial impurities that one can introduce in $\mathcal{N}=2$ theories, but they are not unique. In fact, in Sec. \ref{s:N2}, we explicitly show how to generate an infinite number of them. 



\section{$\mathcal{N}=2$ SUSY BPS impurities - a general construction}
\label{s:N2}

Let us consider a scalar theory in $2+1$ dimensions with a complex field $\phi$. The general BPS equations in terms of the auxiliary fields can be written as follows
\be
F=\pa_{z}\phi e^{i\gamma} \,\, \text{or} \,\, F=\pa_{\bar{z}}\phi e^{i\gamma}\label{sec:N2:BPS}
\ee
where $\gamma$ is an arbitrary constant phase. Now we have two complex SUSY generators (four real supercharges) and the BPS structure is more complex. When $F=0$, for example in the $CP^1$ model, the BPS solutions take the familiar Cauchy-Riemann form and they preserve half of the supersymmetry. This can be seen easily from the SUSY transformation of the fermions
\be
\delta\psi_\alpha =i\sqrt{2}\sigma^{\mu\,\, \dot{\alpha}}_{\,\alpha} \bar{\xi}_{\dot{\alpha}}\pa_\mu\phi+\sqrt{2}\xi_\alpha F \label{sec:N2:fer}
\ee

It is easy to see that for 
\be
\bar{\xi}_{\dot{1}}=i \bar{\xi}_{\dot{2}}\,\, \text{and}\,\, \xi_1=-i\xi_2\label{sec:12:BPS}
\ee
and $\pa_z\phi=0,\,\pa_t\phi=0$ the condition $\delta\psi_\alpha=0$ is satisfied. We say therefore that if $F=0$ the BPS solitons are $1/2$-BPS. If $F\neq 0$ we need an extra constraint, namely 
\be
\xi_2=i \bar{\xi}_{\dot{2}}e^{-i\gamma}.\label{sec:14:BPS}
\ee
 The expression (\ref{sec:N2:fer}) can be rewritten as 
\bea
\delta\psi_1&=&\sqrt{2}\bar{\xi}_{\dot{2}}\left(\pa_0\phi+e^{-i\gamma}F-\pa_z\phi\right),\\
\delta\psi_2&=&-i\sqrt{2}\bar{\xi}_{\dot{2}}\left(\pa_0\phi-e^{-i\gamma}F+\pa_z\phi\right).
\eea
The condition $\delta\psi_\alpha=0$ is again achieved for static solutions satisfying (\ref{sec:N2:BPS}), but this time only preserving $1/4$ of the supersymmetry. 

\subsection{$1/4$ and $1/2$ BPS impurities: scalar sector}

The general commutator between two supersymmetric transformations has the form
\be
[\delta_\eta,\delta_\xi]=-2i \left(\eta \sigma^\mu \bar{\xi}-\xi\sigma^\mu \bar{\eta}\right)\pa_\mu
\ee 
If we restrict the SUSY algebra to the $1/4$ subspace by imposing (\ref{sec:12:BPS}) and (\ref{sec:14:BPS}) we have
\be
[\delta_\eta,\delta_\xi]=8 e^{-i\gamma}\bar{\eta}_{\dot{2}}\bar{\xi}_{\dot{2}}\pa_0.
\ee
Since there is only one SUSY parameter we have
\be
\delta_\eta \delta_\xi=4 e^{-i\gamma}\bar{\eta}_{\dot{2}}\bar{\xi}_{\dot{2}}\pa_0.\label{sec:14:algebra}
\ee
As a consequence, the results found in Sec. 3 for $\mathcal{N}=1$ SUSY apply here: $Y\in \text{Im}\,\delta\Rightarrow \delta_\eta Y\propto \pa_0 X$. Therefore a impurity of the form $\sigma(x)Y$ preserves $1/4$ of SUSY. 
Regarding the $1/2$ preserving impurities in the scalar sector, we do not have a general answer. But since the preservation of $1/2$ SUSY requires $F=0$, the impurity cannot change the BPS equations of the underlying model. For example in the $CP^1$ model one can add the following term
\be
\delta \left( \psi_1 \left(F+\pa_z u\right)\right)\vert_{\gamma=0}=\left(F^2-(\pa_zu)^2\right)+X\pa_0 u+\text{fermions}. \label{sec:12:example}
\ee
Taking into account that (\ref{sec:12:example}) is the image of $\delta$ it preserves at least $1/4$ of SUSY, but since the solution $F=\bar{F}=0$ is still available the BPS impurity solutions are $1/2$ and the BPS equations are not modified.


\subsection{$1/4$ and $1/2$ BPS impurities: gauge sector}

In the gauge sector, the previous results for $1/4$ SUSY still apply since the subalgebra (\ref{sec:14:algebra}) holds. The possible impurities are again in the image of the SUSY transformation. As a consequence, if one considers an impurity of the form 
\be
\sigma(x)\,\delta \left(\lambda_\alpha F(A_\mu, D)\right),
\ee
at least one supercharge will be preserved. Moreover, the SUSY transformation of the fermion in the vector multiplet has the form
\be
\delta \lambda= \sigma^{\mu\nu} \xi F_{\mu\nu}+i \xi D. 
\ee
The reduction to the $1/4$ BPS subspace leads to 
\be
\delta \lambda_2\vert_{\text{static}}= i \xi_2\left(F_{12}-D \right)\label{sec:gauge}
\ee
which is proportional to the BPS equation. Therefore, as in the previous cases, the impurity does not change the BPS bound.  It is also interesting to note that if one considers an impurity of the form (\ref{sec:gauge}) alone, not only $1/4$ but $1/2$ of SUSY is preserved. Unlike in the scalar sector, the gauge $1/2$ BPS impurity also modifies the BPS equation because $D\neq0$ allows for the existence of $1/2$ BPS solitons.


\section{Summary}
\label{sum}

In the present work, we have systematically studied the coupling of an impurity to BPS models in (1+1) and (2+1) dimensions, such that the resulting model preserves half of the BPS property. It means that half of the solitons existing in the original (no impurity) theory still obey the pertinent (impurity dependent) Bogomolnyi equations and saturate a topological energy bound. It turns out that an extremely useful tool for our analysis is provided by supersymmetry. In particular, in (1+1) dimensions, SUSY not only allows to understand both the existence of the impurity BPS models (by relating them to the SUSY transformations of the fermions) and the presence of a generalised translational symmetry in these models \cite{imp-phi4} (relating it to the fermionic zero mode). It also permits to construct the general class of impurity BPS models, see Section \ref{sect-3}. The construction of an impurity BPS model requires a BPS theory without impurity as a starting point, and in (2+1) dimensions not all field theories are BPS. In this case we demonstrated that, whenever a (gauged or ungauged) scalar field theory is BPS, it allows for an impurity BPS generalisation, and SUSY provides an easy way to construct this impurity BPS model. Further, this generalisation is not unique, and each BPS theory allows, in fact, for infinitely many BPS-preserving impurities (see Section \ref{s:N2}). 

So, one important result is that any BPS model in (1+1) and (2+1) dimensions can be extended to a half-BPS impurity model. Since the particular form of the impurity is rather arbitrary, this significantly enlarges the number of theories with the (half) BPS property. In particular, this applies to (2+1) dimensions where, in contrast to the lower dimensional case, the BPS-ness puts very strong restrictions on the form of the action. As a consequence, even in higher dimensions the BPS property is not as rare a feature as one might expect. Owing to the fact that impurities are easily realized in condensed matter, as for example dislocations or defects in a periodic lattice (crystal), we believe that some BPS-impurity models may find applications to realistic physical systems. 

The importance of (near) BPS impurity models is not only related to the simplicity of their mathematical structure. The crucial observation is that they describe topological solitons (kinks, lumps, vortices and baby Skyrmions) which have zero (small) interaction energy with the impurities. This results in an extremely low energy cost for the manipulation of these objects, which might be very interesting, e.g., for the transport and storage of data by topological solitons (like, for instance, magnetic skyrmions).

Furthermore, it is very intriguing that in the case of the planar models ($CP^1$ and Abelian Higgs models) the BPS preserving impurity has a form similar to the Dzyaloshinskii-Moriya energy which may additionally contribute to an experimental realization of this kind of impurities. 

There are many directions in which one may continue our work. First of all, there are other physically important (2+1) dimensional models which enjoy the BPS property. Therefore, one can search for their half-BPS impurity versions. Here one may mention the non-abelian vortices at critical coupling \cite{non-ab vortex}, other gauged planar soliton systems as the gauged BPS baby Skyrme model \cite{g-bBPS}, \cite{CS-bBPS} and the gauged $O(3)$ \cite{O(3)} as well as the conformal magnetic Skyrmions \cite{bazeia-magnetic}.

Secondly, one should understand how the low energy dynamics, approximated by the geodesic motion on the corresponding moduli space, is affected by the inclusion of the half-BPS preserving impurity. This may find some applications also in the study of multi-soliton interactions in non-impurity models, especially in the cases when impurities would be connected to frozen solitons. Here, the simplest case can be given by the $CP^1$ model \cite{leese}, where the soliton-impurity BPS solutions can be found in an exact form, which provides a very good starting point for (even) analytical investigations of the moduli space dynamics. A similar study has already been performed for the half-BPS impurity $\phi^4$ model, where the scattering of the BPS antikink  on the impurity has been shown to be very well described by the motion on moduli space \cite{imp-phi4}.  It would be even more interesting to analyze the existence of solitons and their dynamics in the Abelian Higgs model with half-BPS impurities of both magnetic \cite{vot-imp-dyn} and Higgs types. 

Thirdly, the construction presented here can also be carried over to theories with more than two spatial dimensions. The suspersymmetry algebra is obviously different, but the (2+1) dimensional examples can serve as a guidance for the construction of the bosonic sector. 

Another interesting direction is related to the fact that our BPS
impurity models
possess  (spatially) modulated vacua. They can even be of a periodic
form if a periodic lattice of impurities
is used. Modulated vacua have been recently investigated in Lorentz
invariant theories \cite{bja2} (with higher order
derivative terms). However, such vacua break the supersymmetry
completely. In light of the results presented here,
it is an interesting question whether also in Lorentz invariant theories a fractional
susy preserving modulated vacuum may be found.

Of course, the most exiting direction is to find condensed matter systems which allow for such a specific, BPS preserving, coupling of the impurity. This requires further studies of the relation between the BPS preserving impurity and the Dzyaloshinskii-Moriya interaction observed here. We plan to investigate this issue in a forthcoming paper. 

\hspace*{0.3cm}

{\bf Acknowledgements.}  C.A and A.W.  acknowledge financial support from the Ministry of Education, Culture, and Sports, Spain (Grant No. FPA2017-83814-P), the Xunta de Galicia (Grant No. INCITE09.296.035PR and Conselleria de Educacion), the Spanish Consolider-Ingenio 2010 Programme CPAN (CSD2007-00042), Maria de Maeztu Unit of Excellence MDM-2016-0692, and FEDER.

\appendix
\section{BPS and EL equations of the baby BPS Skyrme model} 
\label{a1}
Here we prove that the Bogomolnyi equation for the the impurity deformed baby BPS Skyrme model implies the full static Euler-Lagrange equations. First of all let us notice that 
\be
\left( \partial_j \frac{\partial}{\partial \xi^a_j} - \frac{\partial}{\partial \xi^a} \right) B_0=0
\ee
Now we apply the Euler-Lagrange operator to the static energy density of the impurity baby BPS Skyrme model
\be
\mathcal{E}= \lambda^2 \pi^4 B_0^2 + \mu^2 \mathcal{U} - 2\pi^2 \lambda \sigma B_0 +2\mu \sigma \sqrt{\mathcal{U}} + \sigma^2
\ee
Hence, 
\bea
\left( \partial_j \frac{\partial}{\partial \xi^a_j} - \frac{\partial}{\partial \xi^a} \right) \mathcal{E} &=& \\
&=& 2\lambda^2 \pi^4 \left( \partial_j B_0 \right) \frac{\partial B_0}{\partial \xi^a_j} - \mu^2 \frac{\partial \mathcal{U}}{\partial\xi^a} -2\pi^2\lambda \partial_j \sigma \frac{\partial B_0}{\partial \xi^a_j}  - 2\mu \sigma \frac{\partial \sqrt{\mathcal{U}}}{ \partial \xi^a} \label{1} \\ 
&=& 2\lambda \mu \pi^2 \left( \partial_j \sqrt{\mathcal{U}} \right) \frac{\partial B_0}{\partial \xi^a_j} - \mu^2 \frac{\partial \mathcal{U}}{\partial\xi^a} - 2\mu \sigma \frac{\partial \sqrt{\mathcal{U}}}{ \partial \xi^a} \\ 
&=& \lambda \mu \pi^2 \frac{1}{\sqrt{\mathcal{U}} }  \frac{\partial \mathcal{U}}{\partial\xi^b} \xi^b_j \frac{\partial B_0}{\partial \xi^a_j} - \mu^2 \frac{\partial \mathcal{U}}{\partial\xi^a} - \mu \sigma \frac{1}{\sqrt{\mathcal{U}} }   \frac{\partial \mathcal{U}}{ \partial \xi^a} \label{3} \\ 
&=& \lambda \mu \pi^2 \frac{1}{\sqrt{\mathcal{U}} } \frac{\partial \mathcal{U}}{\partial\xi^b} \delta^{ab} B_0- \mu^2 \frac{\partial \mathcal{U}}{\partial\xi^a} - \mu \sigma \frac{1}{\sqrt{\mathcal{U}} }   \frac{\partial \mathcal{U}}{ \partial \xi^a}  \\
&=& \left( \lambda   \pi^2  B_0 - \mu \sqrt{\mathcal{U}}  -   \sigma    \right) \frac{\mu}{\sqrt{\mathcal{U}}} \frac{\partial \mathcal{U}}{ \partial \xi^a}=0
\eea 
where in (\ref{1}) we have inserted the Bogomolnyi equation $ \lambda   \pi^2  B_0 = \mu \sqrt{\mathcal{U}} +   \sigma$ while in (\ref{3}) we have used identity $\xi^b_j \frac{\partial B_0}{\partial \xi^a_j}  = \delta^{ab}$.

\end{document}